\documentstyle[12pt]{article}
\author{V.I.Nazaruk\\
Institute for Nuclear Research of RAS, 60th October\\
Anniversary Prospect 7a, 117312 Moscow, Russia\\
(e-mail: nazaruk@inr.msk.su)}
\date{}
\title{$n\bar{n}$ Transitions
}
\begin{document}

\maketitle
\begin{abstract}

We confirm our previous limit on the free-space $n\bar{n}$ oscillation time
$\tau_{min}=4.7\cdot10^{31}y$. The approach is verified by the example of
exactly solvable model. We also discuss the all existing remarks. The
approach with finite time interval includes potential model as an extremely
simplified special case. It is shown that potential model is inapplicable
to the problem in hand because for $\bar{n}$-nucleus decay probability it gives
$W=2$. In particular, as we transparently show, this "decay law" (normalized
by factor 2) is used by Dover, Gal and Richard for refutation of our results.
\end{abstract}

\vspace{5mm}


\vspace{5cm}

INR-995/April 1995

\newpage

\setcounter{equation}{0}

Recently, the calculations of $n\bar{n}$-mixing in nuclear matter beyond
potential model have been done[1]. It was shown that the $n\bar{n}$-
transitions in medium are suppressed only by factor 0.5, as a result of which

\begin{equation}
\label{1}\tau_{n\bar{n}}>\tau_{min}=T/\sqrt{2}=4.7\cdot10^{31}y\,
\end{equation}
where, $T=6.5\cdot10^{31}y$[2]. This lower bound increases the previous one by
31 orders of magnitude. If this limit is true, it would rule out many
existing GUTs. Certainly, this unusual result calls for comprehensive
investigation.

The aim of this paper is: 1.To check the validity of the approach
[1] by the example of the exactly solvable model. 2.To comment upon the main
existing remarks. 3.To find out the conditions such that the restriction (1)
is invalid. In doing so the reasons for disagreement with the previous results
as well as the connection between S-matrix calculations and the approach with
finite time interval are studied.

{\bf 1}. Let us consider the $n\bar{n}$ transitions in nuclear matter (see Fig.
2 of Ref.[1]). The self-consistent neutron potential $U_n$ is included in the
neutron wave function $|n_p\!>$, $p=({\bf p}^{2}_{n}/2m+U_n,{\bf p}_n)$. Block
$T$ is described by Hamiltonian $H(t)$=(all $\bar{n}$-medium interactions)
$-\;U_n$. For the whole process probability $W(t)$ we get[1]

\begin{equation}
\label{2}W(t)=W^a(t)+W^b(t),\;\;\;\;\;W^a(t)=\epsilon^2t^2\,
\end{equation}
\begin{equation}
\label{3}W^b(t)=-\epsilon^2\int_0^tdt_{\alpha}\int_0^{t_{\alpha}}dt_{\beta}%
W_{\bar{n}}(t_{\alpha},t_{\beta}),\;\;\;\;W_{\bar{n}}(\tau)=2ImT_{ii}^%
{\bar{n}}(\tau)\,
\end{equation}
where $W^a$,$W^b$  are the contributions of Fig.2a and 2b of Ref.[1]
respectively, $W_{\bar{n}}(\tau)$ is the probability of $\bar{n}$-nucleus
decay in time $\tau=t_{\alpha}-t_{\beta};\; \epsilon=1/\tau_{n\bar{n}}$.
The matrix element $T_{ii}^{\bar{n}}$ corresponds to $\bar{n}$-nucleus decay:

\begin{equation}
\label{4}iT_{ii}^{\bar{n}}(t_{\alpha},t_{\beta})=\sum_{k=1}^{\infty}(-i)^k
<0\bar{n}_p\mid \int_{t_{\beta}}^{t_{\alpha}}dt_1...\int_{t_{\beta}}^{t_{k-1}}
dt_kH(t_1)...H(t_k)\mid0\bar{n}_p>\,
\end{equation}
where $\mid0\bar{n}_p>$ is the state of medium containing the antineutron with
4-momenta p. In contradistinction to the matrix element of the whole process
(see Fig.2 and Eqs.(8),(9) of Ref.[1]) this equation contains only Hamiltonian
of $\bar{n}$-medium interaction (annihilation) and represents a general
expression for decay matrix. Two step process was reduced to decay
of prepared state which follows the exponential decay law

\begin{equation}
\label{5}W_{\bar{n}}(\tau)=2ImT_{ii}^{\bar{n}}(\tau)=W_{exp}(\tau)=
1-e^{-\Gamma t}\,
\end{equation}
where $\Gamma\sim 100 MeV$ is the annihilation width of $\bar{n}$-nucleus. It
corresponds to all $\bar{n}$-nucleus interactions followed by annihilation.
However, the main contribution gives annihilation without rescattering of
$\bar{n}$[3], because $\sigma_{ann}>2\sigma_{sc}$. From (3) and (5) we get

\begin{equation}
\label{6}W^b(t)=\epsilon^2t^2\left[-\frac{1}{2}+\frac{1}{\Gamma t}+\frac{1}%
{\Gamma^2t^2}\left(e^{-\Gamma t}-1\right)\right].\,
\end{equation}
For our problem  $T=6.5\cdot10^{31}y$[2] and in Eq.(3) one can put
$W_{\bar{n}}(\tau)=1$. Then $W(t)=\epsilon^2t^2/2$. If theoretical calculation
of $T_{ii}^{\bar{n}}(\tau )$ is required, Eq.(5) will serve as a test.

Let us consider the potential block $T$ model, $H=\delta U=U_{\bar{n}}-U_n=
const $, where $U_{\bar{n}}=ReU_{\bar{n}}-i\Gamma /2$ is the $\bar{n}$-nucleus
optical potential. Direct calculation of $T_{ii}^{\bar{n}}(\tau )$ gives

\begin{equation}
\label{7}
\begin{array}{c}
T_{ii}^{\bar{n}}(\tau)=i[1-\exp (-i\delta U\tau )],\\
W_{\bar{n}}^{pot}(\tau)=2ImT_{ii}^{\bar{n}}(\tau)=2\left[1-e^{-
\Gamma \tau/2}\cos (\tau Re\delta U)\right].\,
\end{array}
\end{equation}
Here $W_{\bar{n}}^{pot}$ is the $\bar{n}$-nucleus decay law obtained in the
potential approach framework. Substituting Eq.(7) in Eq.(3), one obtains

\begin{equation}
\label{8}W_{pot}^b(t)=-\epsilon^2t^2\left[1+2Imi/(\delta Ut)^2(\exp (-i\delta
Ut)+i\delta Ut-1)\right].\,
\end{equation}
The whole process probability predicted by potential model is

\begin{equation}
\label{9}W_{pot}=W^a+W^b_{pot}=2Imi(\epsilon/\delta U)^2[1-i\delta Ut-
\exp (-i\delta Ut)].\,
\end{equation}
This expression coincides with the solution of Schrodinger equations:

\begin{equation}
\label{10}(i\partial_t+\nabla^2/2m-U_n)n(x)=\epsilon \bar{n}(x),\;\;\;
(i\partial_t+\nabla^2/2m-U_{\bar{n}})\bar{n}(x)=\epsilon n(x).
\end{equation}
Therefore: (i) The finite time approach was verified by the example
of exactly solvable potential model. It is involved in Eqs.(2),(3) as a
special case. (ii) The self-energy part $\Sigma=\delta U$ have been obtained
dynamically, starting with singular diagrams (see Figs. 3d,3e of Ref.[1]).
Consequently, the potential model substantiation was given (from view
point of singularities, other than physics!\ ).

In the strong absorption region $\Gamma t\sim \mid \delta Ut\mid>>1$ the
potential approach does not correspond to the physics of decays. Really, let
us return to the decay law (7). When $\tau>>1/\Gamma \sim 10^{-24}s$,

\begin{equation}
\label{11}W_{\bar{n}}^{pot}=2.
\end{equation}
The decay matrix $T_{ii}^{\bar{n}}(\tau)$ obtained by means of $U_{\bar{n}}$
gives incorrect result for observable value $W_{\bar{n}}$. Taking into account
that $W$ is represented through $W_{\bar{n}}$ we conclude that {\em potential
model is inapplicable to the problem under study}.

In the region  $\mid \delta Ut\mid<<1$ potential model has a sense:

\begin{equation}
\label{12}W_{\bar{n}}^{pot}(\tau)=W_{exp}(\tau)=\Gamma\tau.
\end{equation}
It remains to see the reason of enormous quantitative disagreement with our
results. The formal answer is that the overall factor 2 in Eq.(7), compared to
Eq.(5), leads to the full cancellation of the $\epsilon^2t^2$ terms in Eq.(9).
The strong result sensitivity was to be expected because when $\delta U=0$
antineutron Green function $G\sim 1/0$ that is a consequence of zero momentum
transfer in the $\epsilon$-vertex. $q=0$ is a peculiar point corresponding to
transition between degenerate states. We will return to this question below.

The standard difinition of decay law is

\begin{equation}
\label{13}W_{\bar{n}}(\tau)=1-\mid U_{ii}\mid ^2,
\end{equation}
where $U$ is evolution operator. {\em Taking into account that unitarity
condition} $UU^+=1$ {\em must be fulfilled for} $U$-{\em matrix of any
process} we immediately get Eq.(3): $W_{\bar{n}}(\tau)=2ImT_{ii}^{\bar{n}}
(\tau)$. {\em We would like to stress that it is sufficient to put}
$W_{\bar{n}}(\tau)=1$. In any case $W_{\bar{n}}\leq 1$ and
$W\geq \epsilon^2t^2/2$.

As for potential model situation is obvious: for $H=\delta U=const$ from Eqs.
(13) and (4) we immediately get Eq.(11).

Summing up we remind that potential model was used only for finite
time approach verification. In the strong absorption region $\Gamma t>>1$ the
potential approach is invalid in principle and so, for realistic problem the
model-independent method based on Eq.(5) has been proposed. Moreover, {\em one
can put} $W_{\bar{n}}=const\leq 1$. In this case {\em only unitarity condition
is employed}.

{\bf 2}. It is very unpleasant to respond to hasty and erroneous criticism,
however it should be made.

Dover, Gal and Richard[4] assert that for realistic problem Eq.(7) must be
used, other than (5). They calculate $W(t)$ substituting Eq.(7) and certainly
get $W=W_{pot}$. On this basis they assert that potential model is correct.
The fact that $W_{\bar{n}}^{pot}=2$ is ignored. By the other words Dover, Gal
and Richard calculate $W$ substituting for $\bar{n}$-nucleus decay
probability  $W_{\bar{n}}=2$.

Dover at al.\  also claim that I am using the equation $W=-2iT_{ii}$. This
expression is applied by them (see Eq.(2) of Ref.[4]) and not by us, because
we know that $W=2ImT_{ii}$. At last I would like to remind that
calculation presented in Ref.[4] have been done in our previous paper (see
last but one paragraph of Ref.[1]), however conclusion was opposite.

The strong absorption region  $\Gamma t\sim \mid \delta Ut\mid>>1$ is
interesting for us. Sometimes it is proposed to eliminate Fig.2a (Eq.(8)) of
Ref.[1]. In this case: (1) For exactly solvable model we have $W_{pot}=W_{pot}
^b=-\epsilon^2t^2<0$. For our calculations $W=W^b=-\epsilon^2t^2/2<0$. (2)
Free-space oscillations $(\rho \rightarrow 0, H \rightarrow 0)$  are not
reproduced. (3) At last, Fig.2a corresponds to the first term of the formal
$U$-matrix expansion and the iterative series for (10).

Krivoruchenko's preprint[5] is completely erroneous. We will cite only two
main points. (1)"...the quantity $<TT^+>$ has no probability meaning, along
with the imaginary part of the $T$-matrix element $-2Im<T>$". (See pg.6 of
Ref.[5].) To our knowledge $|T_{fi}|^2$  has a probability meaning by
definition of $T$-matrix. Krivoruchenko's assertion means that the whole theory
of reactions and decays is incorrect. (2) The initial eq.(11) of Ref.[5] must
describe the $n\bar{n}$-transition, annihilation. However, the l.h.s. of eq.
(11) is free of $\bar{n}$ -nucleus interaction at all. The r.h.s. contains
annihilation width $\Gamma$ and coincides with potential model result. We
would like also to get the result without calculation, but some difficulties
emerge in reaching this goal.

One additional comment is necessary regarding t-dependence of the whole
process probability $W(t)$. Eqs.(2),(3) have been obtained in the lowest order
on $\epsilon$ . The precise distribution $W_{pr}(t)$ which allows for the all
orders on $\epsilon$ is unknown. However, $W$ is the first term of the
expansion of $W_{pr}$ and we can restrict ourselves to a lowest order
$W_{pr}=W$, as it usually is for rare decays.

The protons must be in very early stage of the decay process. Thus the
realistic possibility is considered[6] that the proton has not yet entered
the exponential stage of its decay but is, instead, subject to non-exponential
behavior which is rigorously demanded by quantum theory for sufficiently early
times. At first sight in accordance with (1) for $n\bar{n}$-mixing in nuclear
the similar picture should be expected. In fact the situation is more serious.
We deal with two-step process. In attempting to calculate $\Gamma$ in the
framework of standard S-matrix theory we get $\Gamma\sim 1/0$. So there is no
sense to speak about decay law $\exp (-\Gamma t)$. It is necessary to calculate
$W(t)$ as it was done above.

{\bf 3}. Let us try to render the result invalid. Assume for simplicity that
in the $n\bar{n}$-transition vertex (see Figs.2a,2b of Ref.[1]) the zero
component of 4-momenta $q_0$  is transferred. The antineutron Green function is

\begin{equation}
\label{14}G(x-x_1)=-i\int \frac{d^3p_n}{(2\pi)^3} exp (i{\bf p}_n({\bf x}-
{\bf x}_1)-i(t-t_1)({\bf p}^2_n/2m+U_n+q_0))\theta(t-t_1).
\end{equation}
Instead of Eqs.(2),(3) we get

\begin{equation}
\label{15}W_q(t)=W_c(t)+W_d(t)
\end{equation}

\begin{equation}
\label{16}W_c(t)=2Imi(\epsilon/q_0)^2[1-iq_0t-\exp (-iq_0t)].
\end{equation}

\begin{equation}
\label{17}W_d(t)=2\epsilon^2\int_0^tdt_{\alpha}\int_0^{t_{\alpha}}dt_{\beta}
e^{-i\tau Imq_0}\left[ ReT_{ii}^{\bar{n}}(\tau)\sin (\tau Req_0)-0.5W_{\bar{n}}
(\tau) \cos (\tau Req_0)\right],
\end{equation}
When $q_0 =0$, $W_q(t)=W(t)$. Let us consider Eq.(16). If $q_0=\delta U$ then
(16) coincides with $W_{pot}$. In the region $\mid \delta Ut\mid>>1$

\begin{equation}
\label{18}W_c(t)\rightarrow W_c^{\prime }(t)=(2\epsilon ^2Im 1/\delta U)t.
\end{equation}
The S-matrix calculation[7] in the framework of the potential model ($\bar{n}$
appears in the state with $U_{\bar{n}}\neq U_n$) gives the same result

\begin{equation}
\label{19}W_{exp}=\Gamma_{exp}t=W_c^{\prime }(t).
\end{equation}
Consequently, for nonsingular diagrams and big t finite time approach converts
to the S-matrix theory: $W_c\rightarrow W_c^{\prime }=W_{exp}$. First term of
$W_{exp}$ expansion is reproduced or, what is the same, $\Gamma_{exp}=dW_{exp}
/dt$. This result is obvious and should be considered as a test.

The previous lower bound $\tau_{min}\sim 1y$ is obtained from (18).Therefore,
the energy gap $q_0=\delta U\neq 0$ leads to linear t-dependence (18) and,
consequently, to the process suppression $\tau_{min}\sim 1y$. In our
calculation $q_0 =0$ and $W(t)\sim t^2$.  As a result $W(T)/W_{exp}(T)\sim
\Gamma T\sim 10^{62}$.

Eq.(17) analysis corroborates these conclusions. Really, let us take for
simplicity $ReT_{ii}^{\bar{n}}=Imq_0=0$. Eq. (17) becomes

\begin{equation}
\label{20}
\begin{array}{c}
W_d(t)=\epsilon^2\left[\frac{1}{q_0^2}(\cos (q_0t)-1)+\frac{\Gamma }
{q_0^2+\Gamma^2}t\right] \\
+\epsilon^2 \frac{1}{(q_0^2+\Gamma^2)^2}\left[e^{-\Gamma t}(2q_0\Gamma
\sin (q_0t)+(q_0^2-\Gamma^2) \cos (q_0t))+\Gamma^2-q_0^2 \right].
\end{array}
\end{equation}
When $\Gamma t>>1$, the linear t-dependence $W_d\sim \epsilon^2t$ is reproduced
again. However, if $q_0\rightarrow 0$ then the principal contribution gives
the first term and $W_d\sim \epsilon^2t^2$.

{\em The energy gap} $\delta U\neq 0$ {\em is the reason for dramatic process
suppression},or by the other words the restriction (1) is invalid when
$q_0\neq 0$ in the $n\bar{n}$-transition vertex. $\delta U\neq 0$ can be
realized by means of 3-tail ($q \neq 0$ in the $\epsilon$-vertex), that is the
{\em other} GUT; or effectively, in the framework of the potential approach.
It is {\em invalid in the region} $\Gamma t>>1$.

The qualitative process picture is very simple. Both pre- and post-$n\bar{n}$
-transition spatial wave functions of the system coincide. Thereupon, $\bar{n}$
-nucleus decay takes place, which is described by $\Gamma$ . The magnitude of
$\Gamma $ depends on system wave function. However, the result is $\Gamma$-
independent, because when $T\sim 10^{31}y$, $\Gamma T>>1$ and $W(T)=\epsilon^2
T^2/2$. When $\rho \rightarrow 0$, free-space oscillations are reproduced,
which can be interpreted as a superposition of plane waves. In the slight
absorption region $\Gamma t<(5-10)$ our results coincide with the potential
model ones.

In the next paper the following statements will be proved. 1.The contribution
of the corrections is negligible. 2.All the results are also true for any
nuclear model. We will also draw special attention to the qualitative aspects
of the problem.

We do not see any effect which could change the restriction (1).

\newpage

\end{document}